\documentclass{iopart}
\begin{document}

\title[Analytic expressions for elementary solids]{Analytic expressions for gravitational inner multipole moments
of elementary solids \\ and for the force between two rectangular solids}

\author{E. G. Adelberger, Nathan A. Collins\footnote{Present address: Encina Hall West, Room 100, Stanford University, Stanford, CA, 94305-6044} and C.D. Hoyle}
\address{Department of Physics, University of Washington, 
Seattle, Washington 98195-1560}
\ead{hoyle@npl.washington.edu}
\date{\today}
\begin{abstract}
We give analytic expressions for the gravitational
inner spherical multipole
moments, $q_{lm}$ with $l \leq 5$, for eleven elementary solid shapes. 
These moments, in conjunction with their known rotational and translational
properties, can be used to calculate precisely the moments of complex 
objects that may be assembled from the elementary shapes. We also give
an analytic expression for the gravitational force between two rectangular solids at 
arbitrary separations. These expressions are useful for computing the gravitational properties of
complex instruments, such as those used in equivalence principle tests, and in the 
gravitational balancing of drag-free spacecraft.
\end{abstract}
\submitto{\CQG}
\maketitle
\section{Introduction}
As the precision of gravitational experiments increases it is increasingly necessary to 
compute with high accuracy the gravitational fields of complicated objects. This paper provides
analytic results that have proved useful for this task.

Multipole expansions provide a powerful formalism useful in many 
physical applications.  One such application occurs in equivalence principle tests~\cite{su:94,sm:00}where gravity gradients set a fundamental limit to the 
precision of the results. 
Spurious effects from gravity gradients can
be minimized in a systematic way by nulling the low-order multipole 
moments of the differential accelerometer (typically a torsion 
pendulum) and the attractor. In the case of torsion balances, where 
differential acceleration of the test bodies produces a torque 
on the pendulum, expansions in spherical multipoles are particularly 
useful. Evaluating the multipole moments of 
complicated objects presents a challenge. The moments
can be found by decomposing the objects into 
elementary 3-dimensional shapes whose moments can be calculated
exactly. Because the moments have definite
rotational and translational properties\cite{du:97}, a catalog of 
moments of the elementary solids evaluated in any suitable
coordinate system allows one to ``assemble'' realistic objects and 
compute their multipole moments with high precision. The formalism 
of Ref.~\cite{du:97}, together with the moments cataloged 
here, provides a convenient way to understand the gravitational response 
of torsion balances, including the effects of imperfections 
in the instruments themselves.  This paper gives analytic
expressions for the inner multipole moments of 11 elementary solid 
objects that have proved useful in modeling the gravitational response of actual devices.

Another application of analytic results is found in the gravitational compensation of drag-free spacecraft. For example, the LISA (Laser Interferometer Space Antenna) gravitational wave detector will use three such spacecraft, each containing two inertial bodies that form the end mirrors of interferometer arms. Stringent requirements on the test-body DC acceleration and force noise require accurate modeling of gravitational forces, torques, and gradients produced by the surrounding spacecraft enclosure~\cite{ar:05}. Since much of the disturbing mass distribution is in close proximity to the inertial bodies, the multipole formalism is not valid for this task. The analytic expression for the force between two rectangular solids, derived here, provides an accurate tool for assessing these gravitational disturbances. 
%We could
%not find convienient analytic expressions for the outer moments of these
%shapes (except for the trivial case of the sphere) which are best
%evaluated numerically.
%
\section{Inner moments of the elementary solids}
We present analytic expressions for the gravitational (or Coulomb)
inner moments, $q_{lm}$, 
\begin{displaymath}
q_{lm}=\rho \int r^l Y^{\ast}_{lm} ({\hat r}) d^3r ~,
\end{displaymath}
%defined in Ref.~\cite{ad:90}
of the following elementary shapes: sphere, rectangular solid, 
cylindrical annulus, cone, triangular slab, trapezoidal slab, 
cylindrical section, right tetrahedron, hole in a cylinder, hole in a 
plate, and pyramid. All moments are given with respect to the origin, and are proportional 
to the mass density, $\rho$. 
We list all non-vanishing moments with 
$l \leq 5$; the unlisted $l \leq 5$ moments vanish identically. 
\subsection{Sphere}
The sphere of radius $r$ is centered on the origin. 
\begin{eqnarray*}
\fl     q_{00} = \frac{2}{3} \rho \sqrt{\pi}{r^3}
\end{eqnarray*}
\subsection{Rectangular box}
The box is centered on the origin and is specified by the positive
quantities $x$, $y$, and $z$--its dimensions 
along $\hat{x}$, $\hat{y}$, and $\hat{z}$.
\begin{eqnarray*}
\fl q_{00} = \frac{\rho x y z }{2{\sqrt{\pi }}} \\
\fl q_{20} = -q_{00}\frac{1}{24} \sqrt{5} 
      \left(x^2 + y^2 - 2z^2 \right) \\
\fl q_{22} = q_{00}\frac{1}{8}\sqrt{\frac{5}{6}} 
        \left( x^2 - y^2 \right) \\
\fl q_{40} =q_{00} \frac{1}{640}  \big[ 9x^4 + 10x^2y^2 + 
        9y^4 - 40\left(x^2 + y^2 \right)z^2 
         + 24z^4 \big] \\
\fl q_{42} = -q_{00}\frac{1}{64\sqrt{10}}\left(x^2 - y^2 \right) 
        \left[ 3\left(x^2 + y^2\right) - 10z^2 \right] \\ 
\fl q_{44} =q_{00} \frac{1}{128}\sqrt{\frac{7}{10}} \left( 3x^4 - 
        10x^2y^2 + 3y^4 \right)
\end{eqnarray*}
\subsection{Cylindrical annulus}
\label{cyl}
The cylindrical annulus has its axis of symmetry along $\hat{z}$ 
and is symmetric under reflections through the $\hat{x}$-$\hat{y}$ plane. 
It is specified by inner and outer radii, $r_1$ and $r_2$, and
height $h>0$. It need not be a complete annulus, but spans azimuthal 
angles between $\phi_1$ and $\phi_2$, which are measured with respect
to $\hat{x}$.  We define
$\Delta_{\Phi} \equiv \phi_2 - \phi_1$, 
$\phi \equiv (\phi_1 + \phi_2)/2$,
$\Delta_R \equiv r_2 - r_1$, and $r \equiv (r_1 + r_2)/2$.

\begin{eqnarray*}
\fl q_{00} =  \frac{\rho}{2\sqrt{\pi}} h r \Delta_\Phi \Delta_R\\
\fl q_{11} =  -\frac{\rho}{4\sqrt{6\pi}} e^{-i\phi} h \Delta_R \left(
        \Delta_R^2 + 12r^2 \right) \sin \frac{\Delta_\Phi}{2} \\
\fl q_{20} =  -\frac{\rho}{48} \sqrt{\frac{5}{\pi}} h r \Delta_{\Phi}
        \Delta_R \left(3\Delta_R^2 - 2h^2 + 12r^2 \right) \\
\fl q_{22}  =  \frac{\rho}{16} \sqrt{\frac{15}{2\pi}} e^{-2i\phi} h r 
        \Delta_R \left(\Delta_R^2 + 4r^2 \right) \sin \Delta_\Phi \\
\fl q_{31}  =  \frac{\rho}{960} \sqrt{\frac{7}{3\pi}} e^{-i\phi} h \Delta_R 
        \Big[ 9\Delta_R^4 - 20\Delta_R^2 \left(h^2 - 18r^2 \right)
           - 240r^2 \left(h^2 - 3r^2 \right) \Big] 
        \sin \frac{\Delta_\Phi}{2} \\
\fl q_{33}  =  -\frac{\rho}{192} \sqrt{\frac{7}{5\pi}} e^{-3i\phi} h 
        \Delta_R \big(\Delta_R^4 + 40\Delta_R^2 r^2 
 + 80r^4 \big) \sin \frac
        {3\Delta_\Phi}{2} \\
\fl q_{40}  =  \frac{3\rho}{1280\sqrt{\pi}} h r \Delta_\Phi \Delta_R 
        \Big[ 15\Delta_R^4 - 40\Delta_R^2 \left(h^2 - 5r^2 \right)
 + 8    \left( h^4 - 20h^2r^2 + 30r^4 \right) \Big] \\
\fl q_{42}  =  -\frac{\rho}{128} \sqrt{\frac{5}{2\pi}} e^{-2i\phi} h r 
        \Delta_R \Big[ 3\Delta_R^4 - 24r^2 \left(h^2 - 2r^2 \right)  
        + \Delta_R^2 
        \left( -6h^2 + 40r^2 \right)  \Big] \sin \Delta_\Phi \\
\fl q_{44}  =  \frac{\rho}{512} \sqrt{\frac{35}{2\pi}} e^{-4i\phi} h r 
        \Delta_R \big(3\Delta_R^4 + 40\Delta_R^2r^2 
        + 48r^4 \big) 
        \sin 2\Delta_\Phi \\
\fl q_{51}  =  -\frac{\rho}{3584} \sqrt{\frac{11}{30\pi}} e^{-i\phi} h 
        \Delta_R \Big[ 15\Delta_R^6 - 84\Delta_R^4 \left(h^2 - 15r^2 
        \right) 
          + 672r^2 \left(h^4 - 10h^2 r^2 + 10r^4 \right) \\
            + 56 \Delta_R^2 \left(h^4 - 60h^2 r^2 + 150r^4 \right) \Big] 
         \sin \frac{\Delta_\Phi}{2} \\
\fl q_{53}  =  \frac{\rho}{9216} \sqrt{\frac{11}{35\pi}} e^{-3i\phi} h 
        \Delta_R \Big[ 15\Delta_R^6 - 56\Delta_R^4 h^2 
         + 140\Delta_R^2 \left( 9\Delta_R^2 - 16h^2 \right)r^2 \\ 
        + 560 \left( 15\Delta_R^2 - 8h^2 \right) r^4 
         + 6720r^6 \Big] \sin \frac{3\Delta_\Phi}{2}  \\
\fl q_{55}  =  -\frac{3\rho}{5120} \sqrt{\frac{11}{7\pi}} e^{-5i\phi} h 
        \Delta_R \Big(\Delta_R^6 + 84\Delta_R^4 r^2 
         + 560\Delta_R^2 r^4 + 448r^6 \Big) 
        \sin \frac{5\Delta_\Phi}{2} 
\end{eqnarray*}
\subsection{Cone}
\label{cone}
The (truncated) cone is symmetrical about $\hat{z}$, and is 
specified by upper and lower radii, $r_1$ and $r_2$, and height $h>0$.  
The cone extends a distance $h/2$ above and below the $\hat{x}$-$\hat{y}$ plane.
Complete cones have vanishing values of $r_1$ or $r_2$.
\begin{eqnarray*}
\fl q_{00}  =  \frac{\rho}{6} \sqrt{\pi} h \left( r_1^2 + r_1 r_2 + r_2^2 
        \right)  \\
\fl q_{10}  =  \frac{\rho}{8} \sqrt{\frac{\pi}{3}} {h^2} \left( r_1^2 + 
       2r_1r_2 + 3r_2^2 \right)  \\
\fl q_{20}  =  \frac{\rho}{24} \sqrt{\frac{\pi}{5}} h \Big[ 2h^2 
        \left( r_1^2 + 3r_1r_2 + 6r_2^2 \right) - 3\big( r_1^4 
        + r_1^3r_2 
          + r_1^2 r_2^2 + r_1 r_2^3 + r_2^4 \big) \Big]  \\
\fl q_{30}  =  \frac{\rho}{240} \sqrt{7\pi} {h^2} \Big[ 2h^2 \left( r_1^2 + 
        4r_1r_2 + 10r_2^2 \right) - 3\big( r_1^4 + 2r_1^3r_2 
         + 3r_1^2r_2^2 + 4r_1r_2^3 + 5r_2^4 \big) \Big] \\
\fl q_{40}  =  \frac{\rho}{560} {\sqrt{\pi}} h \Big[ 8h^4 \left( r_1^2 
        + 5r_1r_2 + 15r_2^2 \right) 
         - 12h^2 \left( r_1^4 + 3r_1^3r_2 + 6r_1^2r_2^2 
        + 10r_1r_2^3 + 15r_2^4 \right) \\
          + 15 \left( r_1^6 + r_1^5r_2 + r_1^4r_2^2 + 
        r_1^3r_2^3 + r_1^2r_2^4 + r_1r_2^5 + r_2^6 \right) 
        \Big]  \\
\fl q_{50} =  \frac{\rho}{2688} \sqrt{11\pi} h^2 \Big[ 8h^4 \left( r_1^2 +
      6r_1r_2 + 21r_2^2 \right) 
         - 12h^2 \left( r_1^4 + 4r_1^3r_2 + 
      10r_1^2r_2^2 + 20r_1r_2^3 + 35r_2^4 \right) \\
          + 15 \big( r_1^6 + 2r_1^5r_2 + 3r_1^4r_2^2 + 4r_1^3
      r_2^3 + 5r_1^2r_2^4 + 6r_1r_2^5 
         + 7r_2^6 \big) \Big] 
\end{eqnarray*}
\subsection{Triangular slab}
This shape has reflection symmetry about the $\hat{x}$-$\hat{y}$ plane
and thickness $h>0$. The triangular faces have vertices at 
$(x,y)$ = $(0,0)$, $(d,y_1)$ and $(d,y_2)$.
%, with $d>0$. 
The restriction that one side be parallel to $\hat{y}$ is easily overcome 
using the rotational properties of the moments \cite{du:97}.
\begin{eqnarray*}
\fl q_{00} =  \frac{\rho}{4\sqrt{\pi}} h |d\left( y_1 - y_2 \right)| \\
\fl q_{11}  =   q_{00} 
        \left[- 2d + i\left( y_1 + y_2 \right) \right] \\
\fl q_{20} =  - q_{00}\frac{\sqrt{5}}{12} \left( 3d^2 - h^2 + y_1^2 + y_1y_2 + y_2^2 \right)  \\
\fl q_{22}  =  q_{00} \frac{1}{4}\sqrt{\frac{5}{6}} \Big[ 3d^2 - y_1^2 - y_1y_2 - y_2^2 
         - 3id \left( y_1 + y_2 \right) \Big] \\
\fl q_{31} =  q_{00}\frac{1}{120}\sqrt{\frac{7}{3}}
         \Big\{ 36d^3 - 18i{d^2} 
         \left( y_1 + y_2 \right) 
           + i \left( y_1 + y_2 \right) 
         \left[ 10h^2 - 9\left( y_1^2 + y_2^2 \right) \right] \\
           - 4d \left[ 5h^2 - 3\left( y_1^2 + y_1y_2 
         + y_2^2 \right) \right] \Big\}  \\
\fl q_{33}  =  q_{00}\frac{1}{8} \sqrt{\frac{7}{5}} 
        \left[ \left( 1 + i \right)d + y_1 - iy_2 \right] 
         \times \left[ \left( 1 + i \right)d - iy_1 + y_2 \right]
        \left[ 2 i d +\left( y_1 + y_2 \right) \right]\\
\fl q_{40} =  q_{00}\frac{1}{80} 
        \Big\{ 30d^4 + 3h^4 
         - 10h^2 \left( y_1^2 + y_1y_2 + y_2^2 \right)
         + 6\left( y_1^4 + y_1^3y_2 + y_1^2y_2^2 + 
        y_1y_2^3 + y_2^4 \right) \\
         + 10d^2 \left[ -3h^2 + 2\left( y_1^2 
        + y_1y_2 + y_2^2 \right) \right] \Big\} \\
\fl q_{42} =  -q_{00}\frac{1}{16\sqrt{10}}
        \Big\{ 20d^4 - 15d^2h^2 
         - 20i{d^3} \left( y_1 + y_2 \right)
        + 5h^2 \left( y_1^2 + y_1y_2 + y_2^2 \right) \\
         - 4\left( y_1^4 + 
        y_1^3y_2 + y_1^2y_2^2 + y_1y_2^3 + y_2^4 \right) 
         + 5id \left( y_1 + y_2 \right) \left[ 3h^2 
        - 2\left( y_1^2 + y_2^2 \right)  \right] \Big\}  \\
\fl q_{44} =  q_{00}\frac{1}{8} \sqrt{\frac{7}{10}}\Big[ 5d^4 + y_1^4 + y_1^3y_2 + y_1^2y_2^2 
  + y_1y_2^3 + y_2^4  
   - 10i{d^3} \left( y_1 + y_2 \right) \\
  + 5id \left( y_1 + y_2 \right) \left( y_1^2 + y_2^2 \right) 
   - 10d^2 \left( y_1^2 + y_1y_2 + y_2^2 \right) \Big]  \\
\fl q_{51} =  - q_{00 } \frac{1}{112} \sqrt{\frac{11}{30}} \Big\{ 
        60d^5  - 30i{d^4}( y_1 + y_2 ) 
         + 6 i {d^2} ( y_1 + y_2 ) \left[ 7h^2 - 5( y_1^2 + y_2^2 ) \right] \\
         - 4{d^3} \left[ 21h^2 - 10 \left( y_1^2 +y_1 y_2+ y_2^2 \right) \right] 
         + 2d \big[ 7h^4  - 14h^2 \left( y_1^2 +y_1 y_2 + y_2^2 \right) \\
         + 6  \left( y_1^4 +y_1^3 y_2 + y_1^2 y_2^2 + y_1 y_2^3+ y_2^4 \right) \big] \\
         - i\big[ 7h^4 ( y_1 + y_2 ) 
       - 21h^2 ( y_1^3 +y_1^2 y_2 + y_1 y_2^2 + y_2^3)\\
         + 10\left( y_1^5 + y_1^4 y_2 + y_1^3 y_2^2 + y_1^2 y_2^3 + y_1 y_2^4 + y_2^5 \right) \big] \Big\}  \\
\fl q_{53} =  q_{00}\frac{1}{48} \sqrt{\frac{11}{35}}  \Big\{ 30d^5 
        - 45i{d^4}\left( y_1 + y_2 \right) 
         + 3i{d^2} \left( y_1 + y_2 \right) 
        \left[ 14h^2 - 5\left( y_1^2 + y_2^2 \right) \right] \\
           - 4{d^3}\left[ 7h^2 + 5\left( y_1^2 + y_1y_2 
        + y_2^2 \right) \right] \\
        - i\left( y_1 + y_2 \right) \left[ 7h^2 \left( y_1^2 
        + y_2^2 \right) 
        - 5\left( y_1^4 + y_1^2y_2^2 + y_2^4 \right) \right] \\
         + d\Big[ 28{h^2}\left( y_1^2 + y_1y_2 + y_2^2 \right) 
        - 18\left( y_1^4 + y_1^3y_2 + y_1^2y_2^2 + y_1y_2^3 + 
        y_2^4 \right) \Big] \Big\} \\
\fl q_{55} =  q_{00}\frac{1}{16} \sqrt{\frac{11}{7}} \left[ -2d + i\left( y_1 + y_2 \right) \right] 
        \left[ d^2 - y_1^2 + y_1y_2 - y_2^2 - id\left( y_1 + y_2 \right)
        \right] \\
           \times \left[ 3d^2 - y_1^2 - y_1y_2 - y_2^2
         - 3id\left( y_1 + y_2 \right)  \right] 
\end{eqnarray*}
\subsection{Trapezoidal slab}
The trapezoidal slab has reflection symmetry about the $\hat{x}$-$\hat{y}$
plane and a thickness $t>0$. Two of its sides are parallel to $\hat{y}$ 
with lengths $w_1$ and $w_2$.  It is symmetric about $\hat{x}$, 
i.e., the two parallel sides are centered at $y=0$ with separation $h>0$.
We define $w \equiv (w_1 + w_2)/2$ and $\Delta_w \equiv (w_1 - w_2)/w$.
\begin{eqnarray*}
\fl q_{00} = \frac{\rho}{2\sqrt{\pi}} h t w  \\
\fl q_{11} = q_{00}\frac{1}{4\sqrt{6}} h \left( \Delta_w - 6 
\right) \\
\fl q_{20} = q_{00}\frac{\sqrt{5}}{96} \left[ 4\Delta_w{h^2} - {\Delta_w^2}w^2 
	- 4\left( 4h^2 - 2t^2 + w^2 \right) \right] \\
\fl q_{22} = -q_{00}\frac{1}{32} \sqrt{\frac{5}{6}}
	\left[ 4\Delta_w{h^2} + {\Delta_w^2}w^2 + 4\left( -4{h^2} + w^2 
	\right) \right] \\
\fl q_{31} = -q_{00}\frac{1}{3840} \sqrt{\frac{7}{3}} h 
	\Big[ 3\Delta_w^3 w^2 - 30{\Delta_w^2}w^2 
	 - 120\left( 6h^2 - 4t^2 + w^2 \right) \\
	 + 4\Delta_w \left( 54h^2 - 20t^2 + 15w^2 \right) \Big] \\
\fl q_{33} = q_{00}\frac{1}{256} \sqrt{\frac{7}{5}} h 
	\Big[ -\Delta_w^3 w^2 + 10{\Delta_w^2}w^2 + 4\Delta_w
	 \left( 6h^2 - 5w^2 \right) 
	+ 40\left( -2h^2 + w^2 \right) \Big] \\
\fl q_{40} = q_{00}\frac{1}{10240} 
	\Big\{ -24{\Delta_w^3}w^2h^2 + 9{\Delta_w^4}w^4 
	- 32\Delta_w{h^2} \big( 24h^2 
	 - 20t^2 + 15w^2 \big) \\
	+ 8{\Delta_w^2}w^2 \left( 24h^2 - 20t^2 + 15w^2 \right) 
	 + 16 \big[ 144h^4 + 24t^4 - 40t^2w^2 + 9w^4 \\
	 - 40h^2\left( 4t^2 - w^2 \right) \big]  \Big\}  \\
\fl q_{42} = q_{00}\frac{1}{1024\sqrt{10}} \Big[
	32\Delta_w \left( 8h^4 - 5h^2t^2 \right) 
	+ 3{\Delta_w^4}w^4 \\
	+ 40{\Delta_w^2}w^2 \left( -t^2 + w^2 \right)
	 + 16 \left( -4h^2 + w^2 \right) \left( 12h^2 - 10t^2 
	+ 3w^2 \right) \Big]  \\
\fl q_{44} = q_{00}\frac{1}{2048} \sqrt{\frac{7}{10}}
	\Big[ 24{\Delta_w^3}w^2h^2 + 3{\Delta_w^4}w^4 
	 + \Delta_w \left( -256h^4 + 480h^2w^2 \right) \\
	 + {\Delta_w^2} w^2\left( -192h^2 + 40w^2 \right)
	 + 16\left( 48h^4 - 40h^2w^2 + 3w^4 \right) \Big] \\
\fl q_{51} = q_{00}\frac{1}{57344} \sqrt{\frac{11}{30}} h 
	\Big\{ 3{\Delta_w^5}w^4 - 42{\Delta_w^4}w^4 
	+ 8\Delta_w^3 w^2\big( 26h^2 - 14t^2 \\ 
	+ 21w^2 \big)
	- 224 \big[ 80h^4 + 24t^4 - 20t^2w^2 + 3w^4 \\
	- 20h^2 \left( 6t^2 - w^2 \right) \big] 
	- 112{\Delta_w^2}w^2 \left[ 14h^2 + 5\left( -2t^2 + w^2 \right) 
	\right] \\
	+ 16\Delta_w \big[ 400h^4 - 252h^2 \left( 2t^2 - w^2 
	\right) %\\
	+ 7\left( 8t^4 - 20t^2w^2 + 5w^4 \right) \big] \Big\} \\
\fl q_{53} = -q_{00}\frac{1}{49152} \sqrt{\frac{11}{35}} h 
	\Big\{ -9\Delta_w^5w^4 + 126{\Delta_w^4}w^4 
	+ 112{\Delta_w^2}w^2 \big( 14h^2 \\
	- 20t^2 + 15w^2 \big)
	- 8\Delta_w^3 w^2\left( 26h^2 - 28t^2 + 63w^2 \right) \\
	- 224 \left[ 80h^4 + 40t^2w^2 - 9w^4 
	- 20h^2 \left( 4t^2 + w^2 \right) \right]\\
	 + 16\Delta_w \left[400h^4 - 84h^2 \left( 4t^2 + 3w^2 
	\right) + 35\left( 8t^2w^2 - 3w^4 \right) \right] \Big\} \\
\fl q_{55} = q_{00}\frac{1}{16384} \sqrt{\frac{11}{7}} h
	\Big[ 3\Delta_w^5w^4 - 42{\Delta_w^4}w^4 
	 + {\Delta_w^3}w^2 \left( -208h^2 + 168w^2 \right)\\
	+ 112\Delta_w^2 w^2\left( 14h^2 - 5w^2 \right) 
	 + 16\Delta_w \left( 80h^4 - 252h^2w^2 + 35w^4 \right)\\
	 - 224\left( 16h^4 - 20h^2w^2 + 3w^4 \right) \Big] 
\end{eqnarray*}
\subsection{Cylindrical section}
The cylindrical section has a cross section consisting of the 
space between a chord of a circle (parallel to the $\hat{y}$ axis) and the arc the chord 
subtends.  This object has reflection symmetry about the $\hat{x}$-$\hat{y}$ 
plane and is specified by its thickness, $t>0$, the circle radius, $r$, 
and the distance along the $\hat{x}$ axis, $d>0$, from the center of the circle to the chord.
We define $s\equiv\sqrt{1 - d^2/r^2}$.
\begin{eqnarray*}
\fl q_{00} = \frac{\rho}{2\sqrt{\pi}} t r \left[ 
        - Sd  
        + r\cos^{-1}\left( \frac{d}{r} \right) \right] \\
\fl q_{11} = \frac{\rho}{\sqrt{6\pi}} S t r 
        \left( d^2 - r^2 \right) \\
\fl q_{20} = \frac{\rho}{24} \sqrt{\frac{5}{\pi}} t r
         \Bigg[ Sd
         \left( 2d^2 - t^2 + r^2 \right) 
         + r\left( t^2 - 3r^2 \right)
         \cos^{-1} \left( \frac{d}{r} \right) \Bigg] \\
\fl q_{22} = \frac{\rho}{2} \sqrt{\frac{5}{6\pi}} 
        S d t r \left( r^2 - d^2 \right)\\
\fl q_{31} = \frac{\rho}{60} \sqrt{\frac{7}{3\pi}} 
        S t r \left( r^2 - d^2 \right)
         \left( 6d^2 - 5h^2 + 9r^2 \right) \\
\fl q_{33} = \frac{\rho}{12} t r \sqrt{\frac{7}{5\pi}} 
        S \left( 6d^4 - 7d^2r^2 + r^4 
        \right) \\
\fl q_{40} = \frac{\rho}{160\sqrt{\pi}} t r \Bigg\{ 
        -\Bigg[ Sd \big( 16d^4 - 20d^2t^2 
        + 3t^4 
        + 2\left( 4d^2 - 5t^2 \right)r^2 + 6r^4 \big) \Bigg] \\
           + 3r\left( t^4 - 10t^2r^2 + 10r^4 \right) 
        \cos^{-1} \left( \frac{d}{r} \right) \Bigg\} \\
\fl q_{42} = \frac{\rho}{8\sqrt{10\pi}} S d t r
        \left( d^2 - r^2 \right) \left( 4d^2 - 5t^2 + 6r^2 \right) \\
\fl q_{44} = -\frac{\rho}{8} d t r \sqrt{\frac{7}{10\pi}} 
        S \left( 8d^4 - 11d^2r^2 + 3r^4 
        \right) \\
\fl q_{51} = \frac{\rho}{112} t r \left(d^2 - r^2 \right)
        \sqrt{\frac{11}{30\pi}} S 
        \Big[ 16d^4 - 28d^2t^2 
         + 7t^4 + 6\left( 4d^2 - 7t^2 \right)r^2 
        + 30r^4 \Big] \\
\fl q_{53} = \frac{\rho}{144} t r \left( d^2 - r^2 \right) 
        \sqrt{\frac{11}{35\pi}} S
        \Big[ -48d^4 
        - 14t^2r^2 + 15r^4 
        + 12d^2\left( 7t^2 - 6r^2 \right) \Big] \\
\fl q_{55} = \frac{\rho}{80} t r \left( d^2 - r^2 \right) 
        \sqrt{\frac{11}{7\pi}} S
        \left( 80d^4 - 48d^2r^2 + 3r^4 \right)
\end{eqnarray*}
\subsection{Right tetrahedron}
This shape consists of a tetrahedron having three mutually perpendicular
triangular faces that meet at the origin. The fourth triangular
face is defined by points at coordinates $x$, $y$ and $z$
along the $\hat{x}$, $\hat{y}$, and $\hat{z}$ axes, respectively.
\begin{eqnarray*}
\fl q_{00} =  \frac{\rho x y z }{12\sqrt{\pi}} \\
\fl q_{10} =  q_{00}\frac{\sqrt{3}}{4}z \\
\fl q_{11} =  -q_{00}\frac{3}{4\sqrt{6}}\left(x-i y \right) \\
\fl q_{20} =  -q_{00}\frac{1}{4\sqrt{5}} \left(x^2 + y^2 - 2z^2 \right) \\
\fl q_{21} =  -q_{00}\frac{3}{4\sqrt{30}}z \left(x-i y \right) \\
\fl q_{22} =  q_{00}\frac{3}{4\sqrt{30}} \left(x^2-ixy-y^2 \right) \\
\fl q_{30} =  -q_{00}\frac{\sqrt{7}}{40}z \left(x^2 + y^2 - 2z^2 \right) \\
\fl q_{31} =  q_{00}\frac{1}{80}\sqrt{\frac{7}{3}}
\left[3 x^3-i x^2 y + x{y^2} -3 i {y^3}-4 \left(x-i y \right) z^2 \right] \\
\fl q_{32} =  q_{00}\frac{1}{8}\sqrt{\frac{7}{30}}z
\left(x^2-i x y-y^2 \right) \\
\fl q_{33} =  -q_{00}\frac{1}{16} \sqrt{\frac{7}{5}} \left(x^2-y^2\right) \left(x-iy \right) \\
\fl q_{40} =  q_{00}\frac{3}{280} 
 \left[3 {x^4}+{x^2}{y^2}+3 {y^4}-4 \left(x^2 + y^2 \right)z^2 + 8z^4 \right] \\
\fl q_{41} =  q_{00}\frac{3}{112 \sqrt{5}}z \left[3 x^3 - i{x^2}y -
3 i{y^3}+ 4iy{z^2}+ x\left(y^2-4z^2 \right) \right] \\
\fl q_{42} =  -q_{00}\frac{3}{56 \sqrt{10}}
 \left[2{x^4} -i{x^3}y -ix{y^3} -2{y^4} - 2\left(x^2 -i x y -y^2 \right)z^2 \right] \\
\fl q_{43} =  -q_{00}\frac{3}{16 \sqrt{35}}z  \left(x^2-y^2 \right) \left(x-iy \right) \\
\fl q_{44} =  q_{00}\frac{3}{8 \sqrt{70}} \left(x^4 -i{x^3}y -x^2 
{y^2} +ix{y^3} +y^4 \right) \\
\fl q_{50} =  q_{00}\frac{1}{4486} \sqrt{11}\,z \left[3{x^4} +{x^2}{y^2} 
+3{y^4} -4 \left(x^2+y^2 \right){z^2} +8z^4 \right] \\
\fl q_{51} =  -q_{00}\frac{3}{448} \sqrt{\frac{11}{30}}
 \Big[5{x^5} -i{x^4}y +{x^3}{y^2} -i{x^2}{y^3} +x{y^4} -5i{y^5}- \\
 2 \left(x+iy \right) \left(3{x^2} -4ixy -3{y^2} \right){z^2}+ 8\left(x-i y \right){z^4}\Big] \\
\fl q_{52} =  q_{00}\frac{3}{64} \sqrt{\frac{11}{210}} z 
 \left[-2{x^4} +i{x^3}y +ix{y^3} +2{y^4} +2 \left(x^2-ixy -y^2 \right){z^2} \right] \\
\fl q_{53} =  q_{00}\frac{1}{128} \sqrt{\frac{11}{35}}  
 \Big[5{x^5} -3i{x^4}y -{x^3}{y^2} -i{x^2}{y^3} -3x{y^4} +5i{y^5}  - 
 4 \left(x^2-y^2 \right) \left(x-iy \right) {z^2}\Big]	\\
\fl q_{54} =  q_{00}\frac{3}{64} \sqrt{\frac{11}{70}}z
\left(x^4 -i{x^3}y -{x^2}{y^2} +ix{y^3} +y^4 \right) {z^2} \\
\fl q_{55} =  -q_{00}\frac{3}{128} \sqrt{\frac{11}{7}} \left(x-iy \right) \left[\left(x^2-y^2\right)^2 + x^2 y^2\right] 
\end{eqnarray*}
\subsection{Cylinder hole}
This shape consists of the volume that would be removed by drilling a hole
of radius $r$ into a cylinder of radius $R$. The symmetry axis of the hole
is along $\hat{z}$, and the cylinder has its symmetry axis along $\hat{y}$.
The moments require the hypergeometric function $_2F_1(a,b;c;x)$.
\begin{eqnarray*}
\fl q_{00} = \rho \sqrt{\pi} r^2 R \:{_2F_1}\!\left(-\frac{1}{2}, 
\frac{1}{2}; 2; \frac{r^2}{R^2}\right) \\
\fl q_{20} = -\frac{\rho}{6} \sqrt{5\pi} r^2 R \Big[ \left( r^2 
  - 2R^2 \right)\:{_2F_1}\! \left(-\frac{1}{2}, \frac{1}{2}; 2;
   \frac{r^2}{R^2}\right) 
+ r^2\:{_2F_1}\! \left(-\frac{1}{2}, \frac{3}{2}; 3;
  \frac{r^2}{R^2}\right) \Big] \\ 
\fl q_{22} = \rho \sqrt{\frac{15\pi}{2}} r^2 R \:\:{_2F_1}\! \left(-\frac{1}{2}, 
  \frac{1}{2}; 2; \frac{r^2}{R^2} \right) \\
\fl q_{40} = \frac{\rho}{40} \sqrt{\pi} r R 
 \Big[ \left( 9r^5 - 40r^3R^2 + 24rR^4 \right) \:{_2F_1}\!\left(-\frac{1}{2}, \frac{1}{2};       2; \frac{r^2}{R^2} \right) \\
 + \left( 13r^5 - 32r^3R^2 \right) \:{_2F_1}\!\left(-\frac{1}{2}, 
  \frac{3}{2}; 3; \frac{r^2}{R^2} \right) 
 + 16r^5 \:{_2F_1}\!\left(-\frac{1}{2}, 
  \frac{5}{2}; 4; \frac{r^2}{R^2} \right) \Big]  \\
\fl q_{42} = \frac{\rho}{8} \sqrt{\frac{\pi}{10}} r^4R \Big[ \left( 6r^2 
  - 20R^2 \right) \:{_2F_1}\! \left(-\frac{1}{2}, \frac{1}{2}; 2; 
  \frac{r^2}{R^2} \right) \\
 + 2\left( r^2 + 10R^2 \right) \:{_2F_1}\! \left(-\frac{1}{2}, 
  \frac{3}{2}; 3; \frac{r^2}{R^2} \right) 
 - 13r^2 \:{_2F_1}\! \left(-\frac{1}{2}, 
  \frac{5}{2}; 4; \frac{r^2}{R^2} \right) \Big] \\
\fl q_{44} = \frac{3\rho}{2} \sqrt{\frac{35\pi}{2}} r^2 R 
 \: \: {_2F_1}\! \left(-\frac{1}{2}, \frac{1}{2}; 2; \frac{r^2}{R^2} \right)
\end{eqnarray*}
\subsection{Plate hole}
This shape consists of the volume that would be removed by drilling a hole of
radius $r$ through a parallel-sided plate of thickness $t$. The plate
is centered on the $\hat{x}$-$\hat{y}$ plane. The hole axis, which passes
through the origin, lies in the $\hat{x}$-$\hat{z}$ plane at an angle $\theta$
measured from $\hat{z}$, where $-\pi/2 < \theta < \pi/2$. We define $s\equiv \tan \theta$.
\begin{eqnarray*}
\fl q_{00} &=& \frac{\rho  t r^2 \sqrt{\pi}(1+s^2)}{2} \\
\fl q_{20} &=& q_{00}\frac{\sqrt{5}}{24}\left[3 r^2 (s^2+2) +t^2 (s^2-2)\right] \\
\fl q_{21} &=& -q_{00}\frac{1}{4}\sqrt{\frac{5}{6}} t^2 s \\
\fl q_{22} &=& -q_{00}\frac{1}{8}\sqrt{\frac{5}{6}} s^2 (3 r^2 + t^2) \\
\fl q_{40} &=& -q_{00}\frac{3}{640}
\left[10 r^4 (8+8 s^2+3s^4) +10 t^2 r^2 (3s^4-8) + t^4 \left(8-24 s^2+3 s^4 \right) \right] \\
\fl q_{41} &=& -q_{00}\frac{3}{64 \sqrt{5}} t^2 s 
\left[5 r^2\left(4+3s^2\right)+t^2 \left(3 s^2-4 \right) \right] \\
\fl q_{42} &=& q_{00}\frac{3}{64}\sqrt{\frac{1}{10}} s^2 
\left[10r^4\left(s^2+2\right)+10 r^2 s^2 t^2+t^4\left(s^2-6\right) \right] \\
\fl q_{43} &=& q_{00}\frac{3}{64}\sqrt{\frac{7}{5}}s^3 t^2\left(5r^2+t^2\right) \\
\fl q_{44} &=& -q_{00}\frac{3}{128}\sqrt{\frac{7}{10}}s^4\left(10r^4+10r^2t^2+t^4\right)~,
\end{eqnarray*}
\subsection{Pyramid}
This shape consists of a symmetric pyramid whose base with side lengths $x$ and $y$ lies in the $\hat{x}$-$\hat{y}$ plane
centered about the origin, and whose apex is at $(x,y,z)=(0,0,h)$.
\begin{eqnarray*}
\fl q_{00} &=& \frac{\rho h x y} {6 \sqrt{\pi}} \\
\fl q_{10} &=& q_{00}\frac{\sqrt{3}h}{4} \\
\fl q_{20} &=& q_{00}\frac{4 h^2 -x^2 -y^2}{8 \sqrt{5}} \\
\fl q_{22} &=& q_{00}\sqrt{\frac{3}{10}}\frac{x^2 -y^2}{8} \\
\fl q_{30} &=& q_{00}\frac{\sqrt{7}}{80} h (4 h^2 -x^2 -y^2) \\
\fl q_{32} &=& q_{00}\sqrt{\frac{7}{30}}\frac{h (x^2 - y^2)}{16} \\
\fl q_{40} &=& q_{00}\frac{3}{4480} \left[ 128 h^4 + 9 x^4 + 10 x^2 y^2 +9 y^4 -32 h^2 (x^2 + y^2) \right] \\
\fl q_{42} &=& q_{00}\frac{3}{448 \sqrt{10}}(x^2-y^2) \left[ 8 h^2 -3 (x^2 + y^2)\right] \\
\fl q_{44} &=& q_{00}\frac{3}{128\sqrt{70}}\left[ 3 x^4 - 10 x^2 y^2 + 3 y^2 \right] \\
\fl q_{50} &=& q_{00}\frac{\sqrt{11}}{7168} h 
  \left[128 h^4 +(x^4 +10 x^2 y^2 +9 y^2 -32 h^2 (x^2 +y^2)  \right] \\
\fl q_{52} &=& q_{00}\frac{1}{512}\sqrt{\frac{33}{70}}h (x^2-y^2)\left[8 h^2 - 3 (x^2+y^2) \right] \\
\fl q_{54} &=& q_{00}\frac{3}{1024}\sqrt{\frac{11}{70}} \left( 3x^4 - 10 x^2 y^2 + 3 y^4  \right) ~,
\end{eqnarray*}
\section{Gravitational force between two rectangular solids at arbitrary separations}
In situations where the test masses are in close proximity, the multipole formulation is not valid. In such cases one can use a rectangular mesh to model complex objects. 
We present here an
analytic expression for the force between two parallel-sided rectangular solids at arbitrary separations. Mass 2 has density $\rho_2$ and is centered at the origin with side half-lengths $(a_2,b_2,c_2)$. Mass 1 has density $\rho_2$ and is 
centered at $(\Delta x,\Delta y,\Delta z)$ with side half-lengths $(a_1,b_1,c_1)$.

\begin{equation}
\fl F_z = \!-G \rho_1 \rho_2 \left[I^2_1-I^2_0-I^2_3+I^2_2-y_0\,(I^1_1\!-\!I^1_0)+(y_1-y_0)(I^1_2\!-\!I^1_1)+y_3(I^1_3\!-\!I^1_2)\right],
\label{eq: rectangles}
\end{equation}
where $I^1_{\rm n}$, $I^2_{\rm n}$, and the $y_{\rm i}$ are derived in the Appendix. Similar expressions yield $F_x$ and
$F_y$. Although the complete expressions are quite long they can be evaluated very quickly on modern computers. The precision attainable with our
expressions may be very useful when analyzing the gravitational balancing of drag-free spacecraft, such as those envisioned for LISA\cite{ar:05}. Our expression may also find use in accurately modeling the Newtonian forces between rectangular test masses in future short-range tests of gravity. 
\section*{Appendix: derivation of the force between two rectangular solids}
The z-component of the gravitational force on mass 2 is given by the 6-dimensional integral
\begin{equation*}
\fl F_z = -G \rho_1 \rho_2 \!\!\int_{\Delta x-a_1}^{\Delta x+a_1} \!\!\!\int_{\Delta y-b_1}^{\Delta y+b_1} \!\!\!\int_{\Delta z-c_1}^{\Delta z+c_1}\!\! \!\int_{-a_2}^{a_2} \int_{-b_2}^{b_2} \int_{-c_2}^{c_2} \!\frac{\rmd U}{\rmd(z_1-z_2)}\! \,\rmd z_2 \,\rmd y_2 \,\rmd x_2 \,\rmd z_1 \,\rmd y_1 \,\rmd x_1~, 
\end{equation*}
where $U=\left[(x_1-x_2)^2+(y_1-y_2)^2+(z_1-z_2)^2\right]^{-1/2}$. We transform the coordinates into sums and differences. For example,
\begin{eqnarray*}
z = z_1-z_2 \nonumber \\
z_s = z_1+z_2 \nonumber \\
%\end{eqnarray*}
%and the Jacobian is 
%\begin{equation*}
\rmd z\,\rmd z_s = 2\,\rmd z_1\,\rmd z_2~.
%\end{equation*}
\end{eqnarray*}
%This is equivalent to a rotation of the coordinate axes by 45$^\circ$.
%, the tricky part is to find the new limits. We may 
Integration over $z_s$ is trivial because the function is independent of this variable. The z-integrations give
\begin{eqnarray*}
\fl \int \rmd z_1 \int \rmd z_2\,\frac{\rmd U}{\rmd (z_1-z_2)}&=&\frac{1}{2}\int \rmd z \int \rmd z_s\,\frac{\rmd U}{\rmd z} = \int^{z_3}_{z_2} U \rmd z-\int^{z_1}_{z_0} U \rmd z \\& & \\
&=& F_0(x,y)+F_3(x,y)-F_1(x,y)-F_2(x,y)~,
\end{eqnarray*} 
where
\begin{equation*}
F_{\rm i}(x,y)\equiv \ln \left(z_{\rm i} + {\sqrt{x^2 + y^2 + z_{\rm i}^2}}\right)~.
\end{equation*}
The limits are given by
\begin{eqnarray*}
z_0 &=& \Delta z-c_1-c_2 \\
z_1 &=& \rm{Min}\,(\Delta z+c_1-c_2,\,\Delta z-c_1+c_2) \\
z_2 &=& \rm{Max}\,(\Delta z+c_1-c_2,\,\Delta z-c_1+c_2) \\
z_3 &=& \Delta z+c_1+c_2~.
\end{eqnarray*}
Analogous expressions define the limits for the other coordinates.
The $x$ integrand involves terms such as $F_{\rm i}(x,y)$ and $x\,F_{\rm i}(x,y)$, which lead to
\begin{eqnarray*}
\fl G^1_{\rm ij}(y)\equiv \int^{x_{\rm j}}\!\!\!\rmd x\,F_{\rm i}(x,y)=x_{\rm j}\left[\ln\left(z_{\rm i} + r_{\rm ij}\right)-1\right] + y\left[\tan^{-1}\!\!\left(\frac{x_{\rm j}}{y}\right) - \tan^{-1}\!\!\left(\frac{x_{\rm j} z_{\rm i}}{y\, r_{\rm ij}}\right)\right]\\ \hspace{0.55in}
+z_{\rm i} \ln\left(x_{\rm j} + r_{\rm ij}\right)
\end{eqnarray*}
and
\begin{equation*}
\fl G^2_{\rm ij}(y)\equiv \int^{x_{\rm j}}\!\!\!\rmd x \,x\,F_{\rm i}(x,y)=\frac{1}{4} \left[2\,z_{\rm i}\,r_{\rm ij}-r_{\rm ij}^2+\,2\,\rho_{\rm j}^2 \,\ln \left(z_{\rm i} + r_{\rm ij}\right) \right],
\end{equation*}
where $r_{\rm ij} = \sqrt{x_{\rm j}^2 + y^2 + z_{\rm i}^2}$ and $\rho_{\rm j} = \sqrt{x_{\rm j}^2 + y^2}$.
After performing the trivial integrals over $x_s$ and $y_s$, five of the six integrations are complete. The final step is the integration over $y$, which contains the terms
\begin{eqnarray*}
\fl \int^y \!\!G^1_{\rm ij}(y\,')\,\rmd y\,'=x_{\rm j}\,y\,\left[\ln \left(z_{\rm i} + r_{\rm ij}\right)-\frac{3}{2}\right] + y\,z_{\rm i}\left[\ln \left(x_{\rm j} + r_{\rm ij}\right)-1\right]\\ \\
+\,\frac{y^2}{2}\,\left[\tan^{-1}\!\! \left(\frac{x_{\rm j}}{y}\right) -\tan^{-1}\!\! \left(\frac{x_{\rm j}\,z_{\rm i}}{y\,r_{\rm ij}}\right)\right]+ \frac{z_{\rm i}^2}{2}\,\left[2\,\tan^{-1}\!\! \left(\frac{y}{z_{\rm i}}\right)\right.\\ \\- \left. \tan^{-1}\!\! \left(\frac{x_{\rm j}\,y}{z_{\rm i}\,r_{\rm ij}}\right)\right]
+\,\frac{x_{\rm j}^2}{2}\,\left[\tan^{-1}\!\! \left(\frac{y}{x_{\rm j}}\right)-\tan^{-1}\!\! \left(\frac{y\,z_{\rm i}}{x_{\rm j}\,r_{\rm ij}}\right)\right]\\ 
\\+ x_{\rm j}\,z_{\rm i}\,\ln \left(y + r_{\rm ij}\right)~,
\end{eqnarray*}
\begin{eqnarray*}
\fl \int^y \!\!y\,'\,G^1_{\rm ij}(y\,')\,\rmd y\,'=\frac{1}{12}\left\{-3\,r_{\rm ij}^2\,(x_{\rm j}+z_{\rm i}) - 4\,x_{\rm j}(y^2-\,z_{\rm i}\,r_{\rm ij})+ 6\,y^2\,z_{\rm i}\,\ln (x_{\rm j} + {r_{\rm ij}})\right.\\ \\
+ 4\,y^3\,\left[\tan^{-1}\!\!\left(\frac{x_{\rm j}}{y}\right)- \tan^{-1}\!\! \left(\frac{x_{\rm j}\,z_{\rm i}}{y\,r_{\rm ij}}\right)\right]%\right.\\
%\\
\left.+ 6\,x_{\rm j}\,y^2\,\ln (z_{\rm i} + r_{\rm ij})\right.\\ \\
-2\,z_{\rm i}^3\,\ln \left(\frac{36\,(x_{\rm j}^2 + z_{\rm i}^2)}{x_{\rm j}^4\,z_{\rm i}^6\,(y^2 + z_{\rm i}^2)(x_{\rm j}+r_{\rm ij})}\right)%\right.\\
\left.-2\,x_{\rm j}^3\,\ln \left(\frac{36(x_{\rm j}^2 + z_{\rm i}^2)}{x_{\rm j}^6\,z_{\rm i}^4(r_{\rm ij}+z_{\rm i})}\right)\right\}~,
\end{eqnarray*}
\begin{eqnarray*}
\fl \int^y \!\! G^2_{\rm ij}(y\,')\,\rmd y\,'=\frac{1}{36}\Biggl(6\left\{2\,x_{\rm j}^3\left[\tan^{-1}\!\!\left(\frac{y}{x_{\rm j}}\right)-\tan^{-1}\!\!\left(\frac{y\,z_{\rm i}}{x_{\rm j}\,r_{\rm ij}}\right)\right] + z_{\rm i}\,(3x_{\rm j}^2+z_{\rm i}^2)\ln(y + r_{\rm ij})\right.\Biggr. \\ \\
\left.\Biggl.+y\,(3x_{\rm j}^2 + y^2)\ln(z_{\rm i} + r_{\rm ij})\right\}-y\,(5r_{\rm ij}+16x_{\rm j}^2+4z_{\rm i}^2-12r_{\rm ij}z_{\rm i})\Biggr)~,
\end{eqnarray*}
and
\begin{eqnarray*}
\fl \int^y \!\!y\,'\,G^2_{\rm ij}(y\,')\,\rmd y\,'=\frac{1}{96}\left[8z_{\rm i}^3\,r_{\rm ij}-\left(3\rho_{\rm j}^2+z_{\rm i}^2\right)^2-6\rho_{\rm j}^2z_{\rm i}^2+12\rho_{\rm j}^4\ln (z_{\rm i} + r_{\rm ij})\right]~.
\end{eqnarray*}
Finally, we define
\begin{eqnarray*}
\fl H_{\rm k}(y)\equiv G^2_{\rm k1}(y)-G^2_{\rm k0}(y)-x_0\,(G^1_{\rm k1}(y)-G^1_{\rm k0}(y))+(x_1-x_0)\,(G^1_{\rm k2}(y)-G^1_{\rm k1}(y)) \\ \\
+x_3(G^1_{\rm k3}(y)-G^1_{\rm k2}(y))-G^2_{\rm k3}(y)+G^2_{\rm k2}(y)
\end{eqnarray*}
and
\begin{equation*}
I^1_{\rm n}\equiv\int^{y_{\rm n}} \!\!\rmd y \left[H_0(y)+H_3(y)-H_1(y)-H_2(y)\right]
\end{equation*}
\begin{equation*}
I^2_{\rm n}\equiv \int^{y_{\rm n}} \!\! y\,\rmd y \left[H_0(y)+H_3(y)-H_1(y)-H_2(y)\right]~,
\end{equation*}
which lead to Equation~\ref{eq: rectangles}.
\appendix
\setcounter{section}{1}
\ack
This work was supported in part by National Science Foundation grants
PHY-9602494 and PHY-0355012. CDH would like to thank M.~Armano for carefully 
checking the rectangular solid force calculation.

\end{document}